\documentclass[aps,prl,superscriptaddress,twocolumn]{revtex4}
\usepackage{amsmath,amssymb}
\usepackage{wasysym}
\usepackage{graphics}
\usepackage[pdftex]{graphicx}
\usepackage{epstopdf}
\usepackage{latexsym}
\usepackage{subfigure}
\usepackage{color}
\usepackage{multirow}
\usepackage[]{natbib}

\newcommand{\be}{\begin{equation}}
\newcommand{\ee}{\end{equation}}
\newcommand{\bw}{\begin{widetext}}
\newcommand{\ew}{\end{widetext}}
\newcommand{\bea}{\begin{eqnarray}}
\newcommand{\eea}{\end{eqnarray}}
\newcommand{\ba}{\begin{array}}
\newcommand{\ea}{\end{array}}

\newcommand{\tcr}{\textcolor{red}}

\newcommand{\nn}{\nonumber}
\newcommand{\bs}{\boldsymbol}
\newcommand{\la}{\langle}
\newcommand{\ra}{\rangle}

\begin{document}

\title{RKKY interactions and anomalous Hall effect in metallic rare-earth pyrochlores}
\author{SungBin Lee}
\affiliation{Department of Physics, University of Toronto, Toronto, Ontario M5S 1A7, Canada}
\author{Arun Paramekanti}
\affiliation{Department of Physics, University of Toronto, Toronto, Ontario M5S 1A7, Canada}
\affiliation{Canadian Institute for Advanced Research, Toronto, Ontario, M5G 1Z8, Canada}
\author{Yong Baek Kim}
\affiliation{Department of Physics, University of Toronto, Toronto, Ontario M5S 1A7, Canada}
\affiliation{Canadian Institute for Advanced Research, Toronto, Ontario, M5G 1Z8, Canada}
\affiliation{School of Physics, Korea Institute for Advanced Study, Seoul 130-722, Korea}

\date{\today}
\begin{abstract}
Motivated by experiments on Pr$_2$Ir$_2$O$_7$, we consider metallic pyrochlore systems 
A$_2$B$_2$O$_7$, where the A-sites are occupied by rare-earth local moments
and the B-sites host 5$d$ transition metal ions with itinerant strongly spin-orbit coupled electrons.
Assuming non-Kramers doublets on the A-site, we derive the RKKY interaction between them 
mediated by the B-site itinerant electrons and find extended non-Heisenberg interactions.
Analyzing a simplified model of the RKKY interaction, we uncover a local moment phase with coexisting spiral
Ising-like magnetic dipolar and XY-like quadrupolar ordering. This state breaks time-reversal and lattice symmetries, 
and reconstructs the B-site electronic band structure, producing a Weyl Metallic phase with an intrinsic anomalous 
Hall effect and an undetectably small magnetization. We discuss implications of our results for Pr$_2$Ir$_2$O$_7$.
\end{abstract}
\maketitle

The metallic pyrochlores, A$_2$B$_2$O$_7$, with a rare-earth A-site ion
and a 5$d$ transition metal B-site ion,
lie at the intersection of exciting recent developments in condensed matter physics. The 
rare-earth moments on the A-site pyrochlore sublattice could lead to quantum
spin ice physics \cite{savary2012coulombic,lee2012generic} from Ising anisotropy and 
{\it geometric frustration},
a feature they share with the
well-studied insulating pyrochlore oxides. The strongly {\it spin-orbit coupled} 
5$d$ conduction electrons,  on
the other hand, contain the seeds of
topological phases like Weyl semimetals or topological insulators 
\cite{pesin,yang2010topological,wan2011topological,witczak2012topological,
go2012correlation,wan2012computational}. The interplay of these
two effects could pave the way for new emergent 
phenomena. The pyrochlore iridate Pr$_2$Ir$_2$O$_7$ provides an example of such a 
metallic frustrated system, with a significant Curie-Weiss temperature 
$\theta_{CW} \approx -20$K, much larger than in the insulating compounds Pr$_2$Sn$_2$O$_7$ or Pr$_2$Ti$_2$O$_7$
\cite{yanagishima2001metal,matsuhira2007metal,matsuhira2002low,bramwell2000bulk}. 
For $T \lesssim 1.7$K, even when an applied 
magnetic field along the $\la 111\ra$ direction is switched off, it
exhibits a significant anomalous Hall effect (AHE) which grows upon cooling
 \cite{nakatsuji2006metallic,machida2007unconventional,machida2009time}, 
although, in contrast to Nd$_2$Mo$_2$O$_7$, 
the uniform magnetization is undetectably small over a range of temperatures.\cite{taguchi2001spin} 
This raises the key issue of
mechanisms underlying the AHE \cite{nagaosa2010anomalous} in materials with geometric frustration 
and strong spin-orbit coupling.

%
\begin{figure}[t]
\scalebox{0.6}{\includegraphics{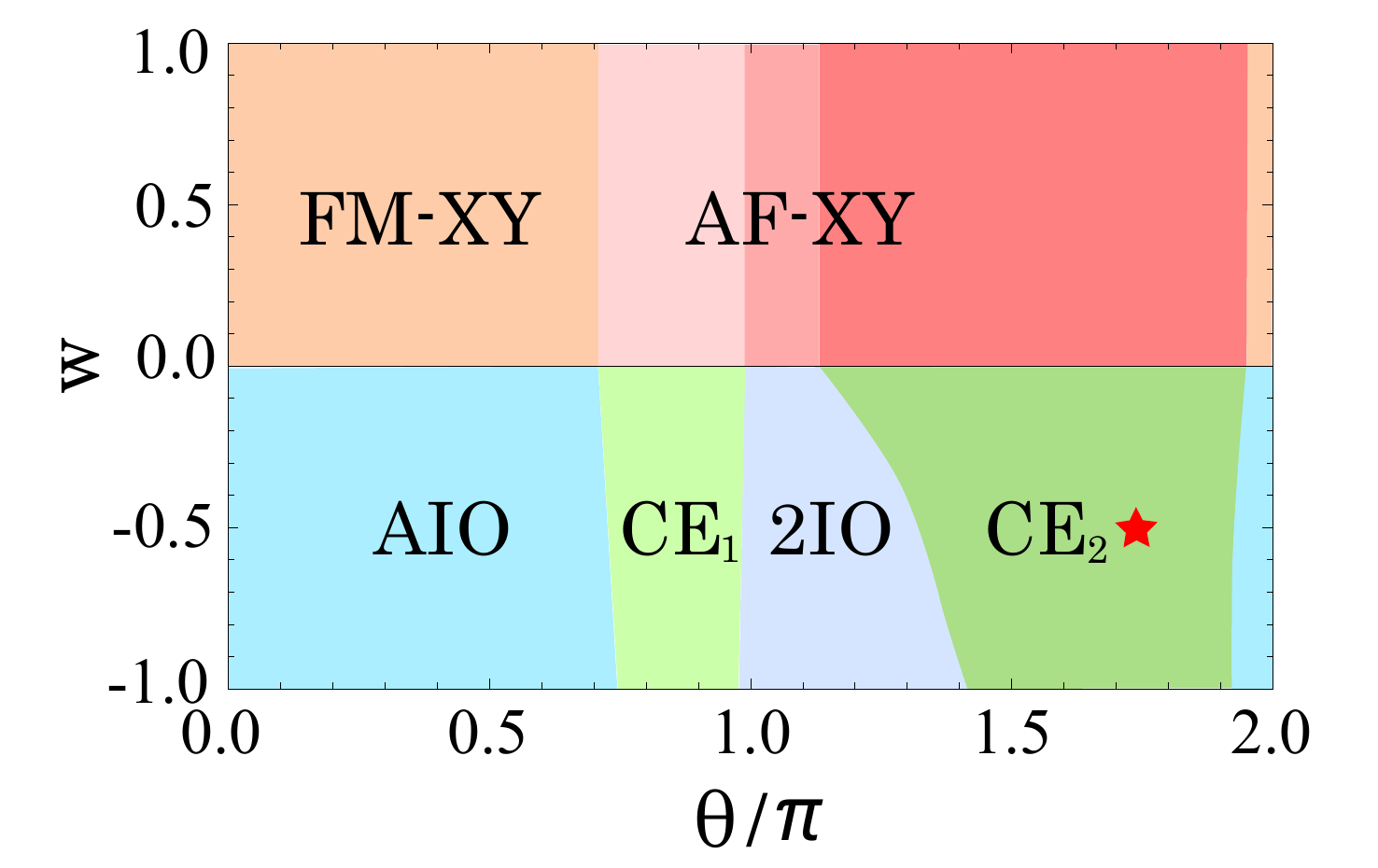}}
\caption{Variational phase diagram of the pseudospin Hamiltonian in Eq.\eqref{eq:2}
relevant to RKKY coupled non-Kramers doublets
as a function of anisotropy $w$ and
$\theta \!=\! \tan^{-1} (J_2/J_1)$.
\underline{$w>0$:} We find  ferro-XY (FM-XY), antiferro-XY (AF-XY), where spins
order in the plane transverse to the local $\la 111 \ra$ axes. These are time-reversal invariant states with 
modulated quadrupolar order.
\underline{$w<0$:} We find Ising all-in/out (AIO), Ising two-in/out (2IO) and states with {\it coexisting} 
modulated Ising-XY orders (CE$_1$,CE$_2$). The CE$_2$ state breaks time-reversal 
and lattice symmetries, producing an anomalous Hall effect for the spin-orbit coupled 
conduction electrons.}
\label{fig:1}
\end{figure}

For metallic pyrochlores, the RKKY interaction between the local $f$-moments 
induced by the conduction electrons is expected to be important. When the $f$-moment is a non-Kramers doublet, such as 
in Pr$^{3+}$, the pseudospin component $\tau^z$ along the local $\la 111 \ra$ direction carries a magnetic dipole 
moment and couples to the conduction electron {\it spin} density, while the $\tau^{x,y}$ components carry a 
quadrupole moment and couple to the electronic {\it charge} density. We show that the resulting RKKY
coupling has a highly non-Heisenberg form and extends beyond the nearest neighbor term.
We propose and study a simplified such
extended XXZ model for non-Kramers doublets using a variational analysis, finding a rich phase diagram, 
shown in Fig.~\ref{fig:1}, which includes incommensurate spiral `coexistence' states (CE$_1$,CE$_2$), with wavevectors 
$(0,q,\pi)$ or $(0,0,q)$, having spatially modulated, magnetic (dipolar) {\it and}
quadrupolar order, i.e., `magneto-quadrupolar  supersolids'. We confirm that such a 
CE$_2$ state also exists within a classical Monte Carlo simulation.
Although the CE$_2$ state has no net local moment magnetization, it reconstructs
the band structure of the spin-orbit coupled 5$d$-electrons, producing Weyl points as well as small Fermi 
pockets, leading to a measurable AHE and an extremely small conduction electron magnetization \footnote{The
presence of Fermi pockets does not allow us to seperate the band and Weyl point contributions to $\sigma_{xy}$}.
We discuss possible implications and predictions for Pr$_2$Ir$_2$O$_7$,
for which our proposal appears to be distinct
from the more widely discussed spin chirality scenario
\cite{kalitsov2009anomalous,machida2009time,udagawa2012,flint2013chiral}.

{\it Basic microscopics and RKKY coupling.---} Since we are motivated by experiments on 
Pr$_2$Ir$_2$O$_7$, 
we briefly review the relevant microscopics of the local moments and conduction electrons in 
this material. The Pr$^{3+}$ ion is in a 4$f^{2}$ configuration, with electron-electron interactions
and strong spin-orbit coupling favoring a total angular momentum state with $J\!\! =\!\! 4$. This nine-fold
degenerate manifold is split by crystal fields with $D_{3d}$
point group symmetry around the Pr$^{3+}$ ion, arising from a cage of
eight O$^{2-}$ ions stretched along the local $\la 111 \ra$ axis. 
The splitting is captured by an effective time-reversal
invariant crystal field Hamiltonian expressed in terms of $J\!\!=\!\!4$ angular momentum operators,
$H_{\rm cef} \! =
\! - \alpha J_z^2 \!+\! \beta [J_z (J_+^3 \!+\! J_-^3) \!+\! {\rm h.c.}] 
\!+\! \gamma (J^6_+ \!+\! J^6_-)$, with $\alpha \!  > \! 0$ and $\beta,\gamma \! \ll \! \alpha$,
leading to a ground state non-Kramers doublet with a dominant $|J_z\!=\!\pm 4\rangle$
component.\cite{onoda2010quantum} Projection to this low energy doublet allows us to define pseudospin-1/2 operators 
$\bs{\tau}$ with $\tau^z \propto J_z$ while $\tau^{\pm} \propto \{J_{\pm} ,J_z\}$. 
Under time-reversal, $\tau^z \! \to \! -\tau^z$,
transforming as the magnetic dipole moment, while $\tau^{x,y}$ are left invariant, transforming
like a quadupole moment.

For Ir$^{4+}$ electrons, strong spin-orbit coupling in the $t_{2g}$ manifold leads to a half-filled effective
$j\!\!=\!\! 1/2$ band. The metallic character of Pr$_2$Ir$_2$O$_7$ suggests that
a tight binding model which ignores strong electronic correlations would be an adequate 
starting point,
\be
H_{\rm{tb} } = \sum_{ij} \sum_{\alpha \beta} c^\dagger_{i \alpha }(t^{\vphantom\dagger}_{ij} \delta^{\vphantom\dagger}_{\alpha \beta} 
+ i\bs{v}^{\vphantom\dagger}_{\ij} \cdot \bs{\sigma}^{\vphantom\dagger}_{\alpha \beta}  ) c^{\vphantom\dagger}_{j \beta},
\label{eq:1}
\ee
where $c^\dagger_{i \alpha} ( c^{\vphantom\dagger}_{ j \beta} )$ denotes the electron 
creation (annihilation) operator at site $i$ with the Kramers pseudospin index $\alpha$ corresponding to 
$j_z\!=\!\pm 1/2$, and ${\bs \sigma} = ( \sigma^x , \sigma^y, \sigma^z)$ are the Pauli matrices. 
We assume $t_{ij}=t$ $(t')$ for nearest (next-nearest) neighbors, and
${\bs v}_{ij} \neq 0$ only for nearest-neighbors and is constrained by lattice symmetries.
\cite{lee2012magnetic}

How does the A-site non-Kramers doublet couple to the B-site conduction electrons? 
Unlike the usual Kondo coupling to a magnetic Kramers doublet, 
time-reversal invariance dictates that the Ising component of the A-site 
pseudospin $\tau_j^z$ at site $j$ (which points along the local $\la 111 \ra$ axis)
couples to {\it spin} density  $j_i^\mu = c_{i \alpha}^\dagger \frac{\sigma_{\alpha \beta}^\mu}{2} c_{i \beta}$
of electrons on the six neighboring B-sites,
while the planar components of the pseudospin $\tau_j^\pm$ couple to the {\it charge} density 
$n_i = c^\dagger_{i \alpha} c_{i \alpha}$ on the neighboring B-sites. 
Keeping $j_i^\mu \tau^z_j$ and $n_i \tau_j^\pm$ terms, 
we find the symmetry allowed Kondo coupling $H_{AB}$ 
with three parameters $c_1,c_2,c_3$. (See Supplementary Materials for details.) 
Integrating out the conduction electrons, we find that
the resulting RKKY interaction has two important features which are insensitive to the details of the Ir
band structure: (i) it allows for significant couplings beyond the nearest-neighbor interaction,
but is negligible beyond the third neighbor; (ii) it is highly anisotropic in spin-space since
$\tau^{x,y}$ and $\tau^z$ interact very differently with the spin-orbit coupled conduction band. 
A complete microscopic set of Kondo couplings should include 
$\tau^z_j 
c_{i \alpha}^\dagger \frac{\sigma_{\alpha \beta}^\mu}{2} c_{\ell \beta}$ or $\tau^{\pm}_j 
c^\dagger_{i \alpha} c_{\ell \alpha}$ with $i \neq \ell$, terms which we have omitted,
and the full characterization may require multiple parameters.
Rather than dealing with such a complex model, we would like to focus on
general and robust features of the RKKY interactions. 
We therefore turn to a study of a simplified RKKY 
Hamiltonian which retains the two key features described above.

{\it Local Moment Model and Phase Diagram. ---}
Motivated by our observation that the RKKY interaction between local moments 
is highly anisotropic in spin space, and has beyond nearest-neighbor terms, 
we study a simplified model in the basis with a 
{\it sublattice-dependent} quantization axis along the {\it local} $\la 111 \ra$ direction,
\bea
H \!\!&=&\!\! -\!\! \sum_{\bs{r},\bs{r'}} J_{{\bs r},{\bs r'}} \Big[ (1\!-\! w)  \tau_{s}^z (\bs{r})  \tau_{s'}^z (\bs{r}')
\!+\! \bs{\tau}^\perp_{s} (\bs{r}) \!\cdot\! \bs{\tau}^\perp_{s'} ( \bs{r'}) \Big].
\label{eq:2}
\eea
Here $J_{\bs{r},\bs{r'}} = J_1 (J_2)$ for nearest (next-nearest) neighbor sites,
$\tau_{s}^z (\bs{r})$ is $z$-component of the pseudospin on sublattice $s$ at site $\bs{r}$, 
$\bs{\tau}^\perp_s({\bs r})$ denotes
the transverse component of the pseudospin which lies in the local XY plane, and $w$ quantifies 
the exchange anisotropy. We set $J_1=J\cos\theta$ and $J_2=J\sin\theta$, and explore the phase 
diagram of this model as a function of $(\theta,w)$.

Treating the spins as classical unit vectors, we minimize the energy using a variational ansatz
\bea
\bs{\tau}_s (\bs{r})  \!\!&=&\!\! {d}_s \hat{\bs{e}}_3 \!+\!
\sqrt{1 - {d}_s^2}~{\cal R}e [ ( \hat{\bs{e}}_1 + i \hat{\bs{e}}_2) e^{i  (\bs{Q} \cdot \bs{r} + \varphi_s) }],
 \label{eq:3} 
\eea
where $\hat{e}_{1,2,3}$ form a triad of orthonormal vectors, so that $|\bs{\tau}_s(\bs{r})|=1$. In the
local coordinate system, this 
ansatz allows, (i) for $d_s=0$, a coplanar spiral with wavevector $\bs{Q}$ with
spins in the $(\hat{e}_1,\hat{e}_2)$ plane, and, (ii) for $d^2_s=1$, a collinear state 
with spins along $\pm \hat{e}_3$. In the isotropic limit, $w=0$, this ansatz recovers
$(0,0,q)$ spirals \cite{lapa2012ground}, while in the Ising limit it allows for 2-in 2-out or 
all-in all-out states. The complete ground state phase diagram from this variational analysis 
is shown in Fig.~\ref{fig:1}. 

For $w>0$, we find states where the spins lie in the local XY plane, forming
phases like ferromagnetic XY, antiferromagnetic XY, or degenerate XY versions of 2-in 2-out states. These
XY ordered states for the non-Kramers ions do not break time-reversal symmetry but
correspond to modulated quadrupolar orders, with no net quadrupole moment and hence
may be termed `antiferroquadrupolar' (AFQ) states. With increasing $\theta$, the different colors in the AF-XY 
region represent states with distinct ordering wavevectors $\bs{Q}=(0q \pi ),(0,0,0), (00q)$.

For $w < 0$, by contrast, we find in addition to well known states like the
ferromagnetic Ising (all-in all-out), and Ising spin ice (2-in 2-out), large parameter regimes which support
coexistence phases (CE$_1$,CE$_2$) with coplanar order involving spatially modulated Ising (magnetic
dipolar) and XY (quadrupolar) order. The CE$_1$ and CE$_2$ states order at the wavevectors $(0,q,\pi)$ and $(0,0,q)$
respectively (or their symmetry related momenta).
Such CE states are `magneto-quadrupolar' supersolids which break time-reversal symmetry 
(defined by $\tau^z \to -\tau^z$), and most lattice symmetries. Below, we focus on the remarkable
physical properties including the AHE of the CE$_2$ state which is (i) robustly present in an unbiased numerical 
energy minimization using a simulated annealing approach
\footnote{Although we have not
mapped out the entire parameter space, we find that the variational ansatz captures most of the phases found in the
simulated annealing approach. Our ansatz does not capture regimes where multiple-$\bs{Q}$
orders are favored - such states are well known to occur in the isotropic case $w=0$ in the regime where we find
CE$_1$ order, and will be discussed separately in future work for $w \neq 0$}, 
and (ii) stable to the addition of weak perturbations such as $(\tau_i^+\tau_j^+ + \tau_i^-\tau_j^-)$
to the simple Hamiltonian in Eq.~\ref{eq:2}.  A more complete study including such terms which break the
local U(1) symmetry will be discussed elsewhere \cite{lee:J1-J2-pyrochlores}.

{\it AHE and Magnetization in the CE$_2$ state.---}
Once the CE$_2$ state is stabilized on the A-site, it imprints effective spatially varying
magnetic fields and chemical potentials on the B-site conduction electrons due to the 
Kondo-type coupling between them. For an ordering wave vector $\bs{Q}$, this mixes
electrons with wave vectors $\bs{k}$ and $\bs{k +Q}$, leading to a Hamiltonian of the form
\bea
H_{\rm{tb}}^B (\bs{k}) = 
\left(
\begin{array}{cc}
 H_{\rm{tb}} (\bs{k}) & H_{AB} (\bs{Q}) 
 \\
 H_{AB} (- \bs{Q} ) &   H_{\rm{tb}} (\bs{k+Q}) 
 \end{array}
\right)
\label{eq:4}
\eea
For the CE$_2$ state in our anisotropic $J_1$-$J_2$ pseudospin model, $\bs{Q}=(00q)$, 
and time-reversal symmetry and all lattice symmetries 
except $C_{2z}$ ($\pi$ rotation along $\hat{z}$ direction) are broken.
Once these broken symmetries are inherited by the resulting
5$d$ electron band structure, one can argue for an intrinsic AHE. The
$C_{2z}$ symmetry leads to $\sigma_{yz}\!=\! \sigma_{xz}\! =\! 0$ since the current
operator $J_z$ is invariant under $C_{2z}$ rotation $(x,y,z) \!\rightarrow\! (-x,-y ,z)$ while
the currents $J_{y}$ and $J_x$ change sign. However,
$\sigma_{xy}$ remains unchanged under this rotation and can thus be nonzero.

\begin{figure}[t]
\scalebox{0.4}{\includegraphics{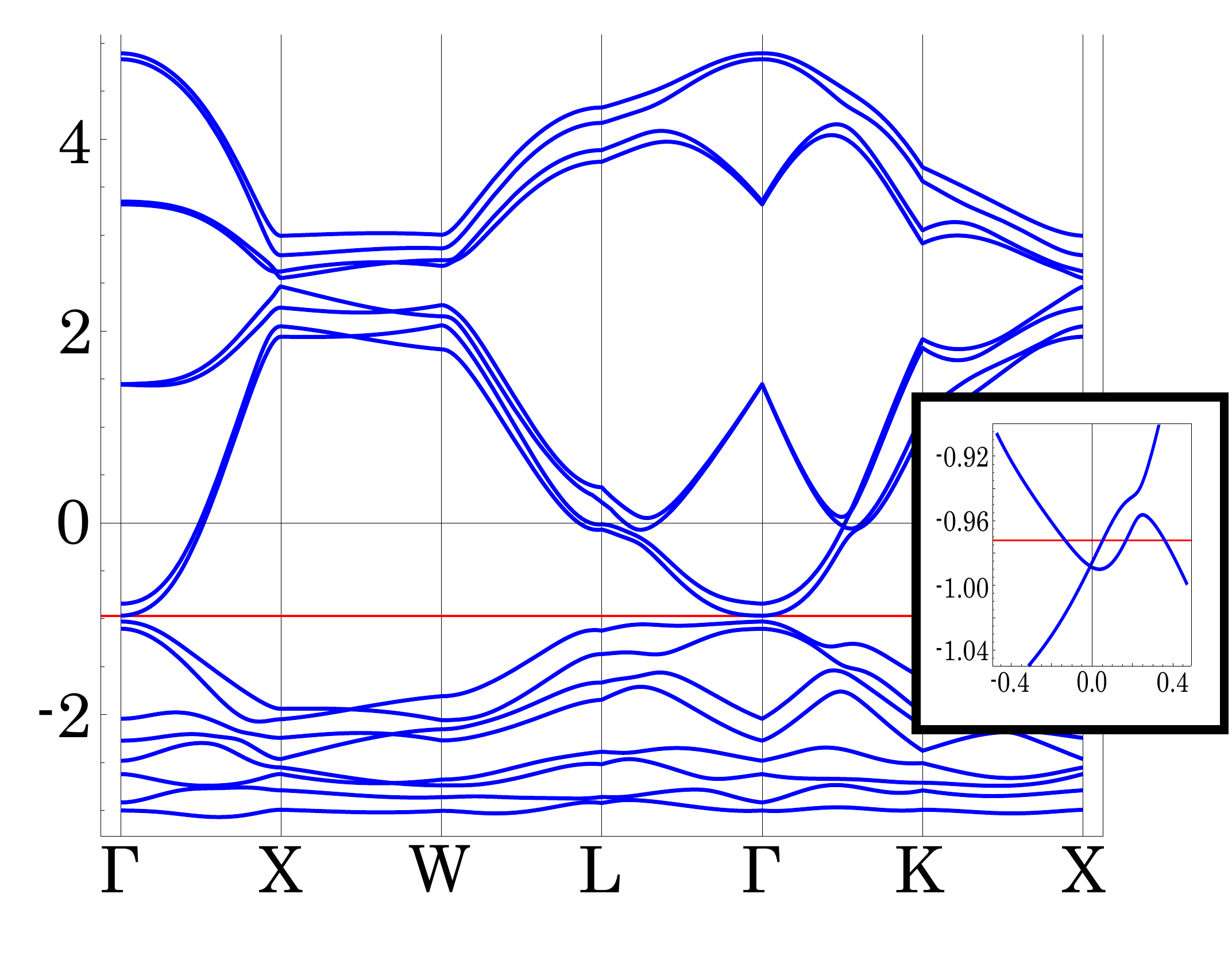}}
\caption{Reconstructed 5$d$ conduction electron band structure (based on Eq.\eqref{eq:4} 
with the energy units of $t$) 
in the presence of Kondo coupling constant $c/t=0.1$ and
the A-site incommensurate CE$_2$ local moment order with wave vector $\bs{Q} \approx 1.2\pi(001)$.
The inset shows one Weyl point and Fermi pockets near $\bs{k} \approx (0,0.12\pi,0.02\pi)$ 
(the horizontal axis represents  $k$ along $\bs{k} = (k,k+0.12\pi,0.02\pi)$). 
Its pair is located at $\bs{k} \approx (0.06\pi, 0.18\pi,0)$. Such a pair of Weyl points is near $\Gamma$ point 
and there is another pair near $\Gamma +\bs{Q}$. Furthermore, additional two pairs exist at 
their $C_{2z}$ symmetry related points. Overall, there are four pairs of Weyl points. 
}
\label{fig:2}
\end{figure}

To explicitly compute the Hall conductivity, we assume the tight-binding parameterization for
the Ir 5$d$ electrons in Eq.\eqref{eq:1}, as relevant to Pr$_2$Ir$_2$O$_7$, choosing
$t = 1$, $t'=-0.1$, and $|\bs{v}_{ij}| =0.2$. These values are
close to those determined from the Slater-Koster parameters.\cite{lee2012magnetic}
Fig.\ref{fig:2} shows an example of the reconstructed B-site band structure 
based on Eq.\eqref{eq:4}. (Red line indicates the Fermi level.) 
We consider the A-site 
incommensurate CE$_2$ local moment order with wave vector 
$\bs{Q} \approx 1.2\pi(001)$, which is appropriate for 
$\theta = 1.72\pi, w= -0.5$ in Eq.\eqref{eq:2} ($\tcr{\star}$ marked in Fig.\ref{fig:1}), 
and the Kondo coupling constant $c/t=0.1$ where $c_1=0$, $c_2=c_3=c$ in $H_{AB} (\bs{Q})$. 
(See Supplementary material)   
The reconstructed band structure generates both Weyl points and Fermi pockets near
$\Gamma, \Gamma+\bs{Q}$. 
There are four pairs of Weyl points in total (two pairs near $\Gamma, \Gamma+\bs{Q}$ 
and other two pairs at their $C_{2z}$ symmetry related points) and 
the inset of Fig.\ref{fig:2} shows 
one of those Weyl points and Fermi pockets near $\bs{k} \approx (0,0.12\pi,0.02\pi)$. 

\begin{figure}[]
\scalebox{0.68}{\includegraphics{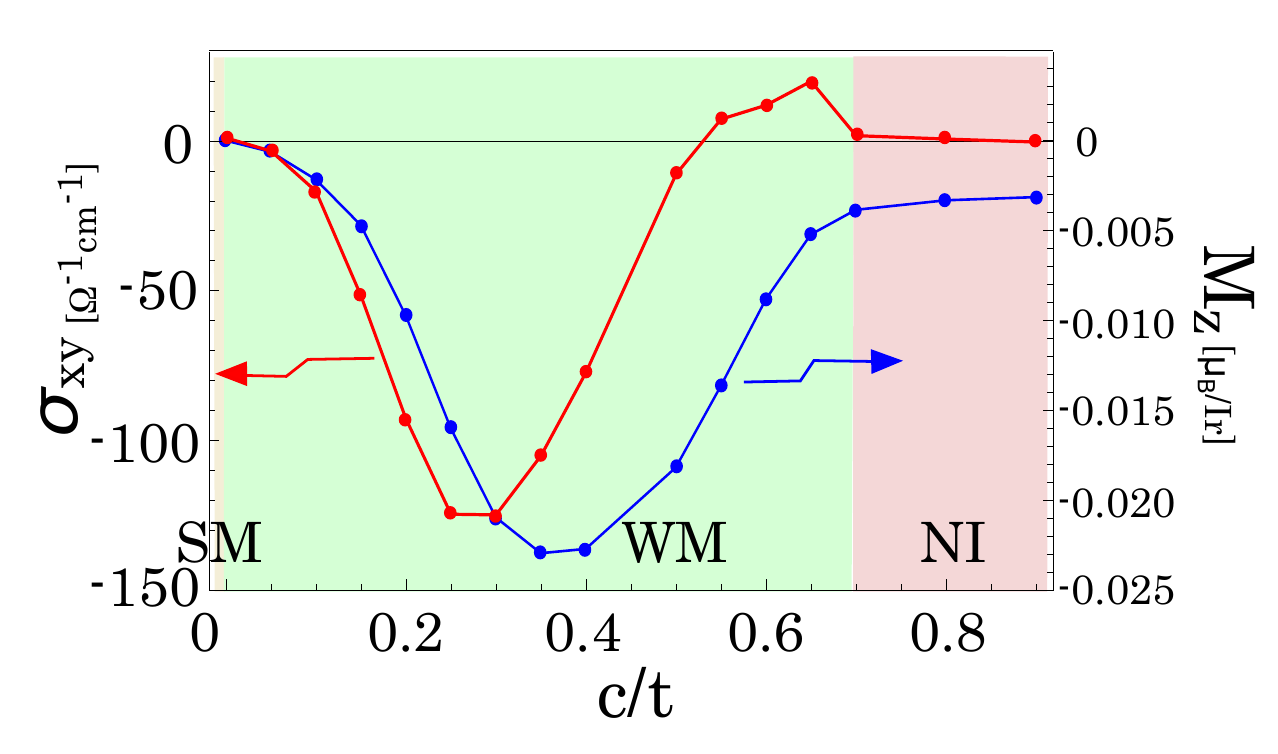}}
\caption{Hall conductivity $\sigma_{xy}$ and magnetization $M_z$ of 5$d$ conduction electrons as a function of 
Kondo coupling constant $c$ based on the reconstructed band structure. (See Fig.\ref{fig:2} for the reconstructed
band structure at $c/t=0.1$)
 For $c=0$, the conduction band is in a semi-metal phase (SM) with 
a quadratic band touching at the $\Gamma$ point. For $0< c/t \lesssim 0.7$, the CE$_2$ state reconstructs the bands 
generating both Fermi pockets and pairs of Weyl points, resulting in a Weyl-metallic phase (WM), with a significant
nonzero $\sigma_{xy}$ which has contributions from Weyl points as well as Fermi surfaces. For $c/t \gtrsim 0.7$, 
the conduction electrons form a gapped normal insulating phase (NI). The magnetization $M_z$ has the same
sign as $\sigma_{xy}$ over a wide range of couplings $c$, although it is extremely small in magnitude, and is nonzero
even in the NI.}
\label{fig:3}
\end{figure}

Fig.\ref{fig:3} shows the explicitly calculated Hall conductivity using the
Kubo formula
as a function of Kondo coupling constant $c$, 
based on the reconstructed B-site band structures. 
The strength of Kondo coupling $c$ determines the magnitude
of the AHE response. $\sigma_{xy}$ initially increases with increasing $c$, 
acquiring contributions from
small electron-like and hole-like Fermi pockets as well as four pairs of Weyl points induced 
by the CE$_2$ order. For $c/t \gtrsim 0.7$, a band gap opens up and $\sigma_{xy}= 0$.

%
%

On symmetry grounds, the AHE must be accompanied by a nonzero uniform magnetization, with $M_x\!=\!M_y\!=\!0$
but $M_z \!\neq \!0$. Remarkably, although the CE$_2$ state has no net magnetization from the local moments,
we find that it induces a nonzero magnetization $M_z$ for the 5$d$ electrons, 
where $M_\mu = \frac{1}{N} \la \sum_{i} \sum_{\alpha \beta} c^\dagger_{i \alpha }  \frac{\sigma_{\alpha \beta}^\mu}{2} c_{i \beta} \ra$.
For small $c$, the net magnetization $M_z$ gets larger, proportional to the density of states (DOS) near the Fermi level, with
the {\it same} sign and a trend qualitatively similar to $\sigma_{xy}$. For large $c/t \gtrsim 0.7$, we find that $M_z \neq 0$ 
although $\sigma_{xy}=0$, signalling a magnetized band insulator.

This dichotomy of a large $\sigma_{xy}$ but a small $M$ can be argued for as follows.
On dimensional grounds, the AHE signal $\sigma_{xy} \sim \frac{e^2}{h} \Delta k$, where the momentum scale
$\Delta k$ must be
induced by an effective `internal magnetic field' $B^{\rm int}$ due to the spontaneous breaking of time-reversal 
symmetry\cite{burkov2011weyl,PhysRevB.84.075129}.
If the $\tau^{x,y}$ order in the CE$_2$ state reconstructs the band structure to produce a
small Fermi pocket, with a Fermi wavevector $k_F$ and an effective mass $m^*$, we expect $\Delta k \sim B^{\rm int} m^* /k_F$
resulting in a large $\sigma_{xy}$ due to a small $k_F$ Fermi pocket, while the magnetization $M \sim B^{\rm int} m^* k_F$ stays 
small. The Weyl point contribution from splitting a quadratic band touching, with a curvature $m^{**}$, on the other hand
leads to $\Delta k \sim \sqrt{m^{**} B^{\rm int}}$\cite{moon2012non}, and a magnetization $M \sim B^{\rm int}$, so that a small $B^{\rm int}$
again leads to a large $\sigma_{xy}$ and a small $M$. The presence of both contributions would lead to a non-linear 
$\sigma_{xy}(M)$.

{\em Application to Pr$_2$Ir$_2$O$_7$: } The CE$_2$ state in our simple model of RKKY coupled non-Kramers ions 
and spin-orbit coupled conduction electrons captures a key
aspect of the experimental data on Pr$_2$Ir$_2$O$_7$: an AHE accompanied by a negligible magnetization.
However, since our ordering wavevector is along the $\la 001 \ra$ direction, the
spontaneous AHE is produced in the $xy$-plane, whereas the spontaneous AHE in Pr$_2$Ir$_2$O$_7$ is seen
for fields along the $\la111 \ra$ direction. This discrepancy may be resolved if such coexisting spiral order had a wavevector
along the $\la 111 \ra$ direction, or if our CE order gave way to a multimode spiral formed by superposing $(0,0,q),(0,q,0),(q,0,0)$
spirals while preserving $C_{3}$ rotation along $\la 111 \ra $ direction but breaking all other lattice symmetries.
The Hall conductivity in that case will naturally be in the plane perpendicular to ${\la 111 \ra}$, in agreement with experiment.
We estimate the magnitude of the AHE (at $c/t=0.1$, which is $c\approx 5\rm{meV}$) in our CE$_2$ state 
to be $\sigma_{xy} \approx -15(\Omega^{-1} \rm{cm}^{-1})$ and $M_z \approx -0.002 (\mu_B /\rm{Ir})$. This is
consistent with the magnitude of the experimentally measured Hall signal, and the absence of any measurable magnetization
over a range of temperatures. At lower temperatures, experiments detect a nonzero remnant magnetization, which may 
arise in our model from the feedback of the magnetized conduction electrons on the local moments, an effect 
we have not taken into account.

The Curie-Weiss temperature is sensitive to $J_2/J_1$, changing sign as a function 
of $\theta$ even within the CE$_2$ state. For couplings $c/t \sim 0.1$, 
we can obtain $\theta_{CW}$ to be 
of the right sign and magnitude, $\theta_{CW} \sim -20K$ as observed experimentally, for parameter
values closer to the all-in all-out phase boundary.

The CE$_2$ state should exhibit two distinct thermal transitions, associated with the onset of Ising and
XY orders, with this splitting being smaller for a weaker anisotropy $w$. However, this simple expectation gets
confounded by two issues. (i) Terms which we have omitted, such as $(\tau_i^+\tau_j^+ + \tau_i^-\tau_j^-)$, 
will break the (staggered) U(1) invariance of the XY terms, and modify the 3d-XY universality class 
of the quadrupolar ordering transition. (ii) Oxygen vacancy defects will produce strong
electric fields which break the $D_{3d}$ point group symmetry around the Pr$^{3+}$ ion, splitting 
the non-Kramers doublet due to extra random terms $\Delta H_{\rm cef} \propto \{J_z,J_\pm\} \sim \tau^{\pm}$ 
in the effective
crystal field Hamiltonian. This would lead to a time-reversal invariant strong random field on $\tau^{x,y}$. 
These effects might conspire smear or destroy the XY ordering transition. However, the time-reversal symmetry 
breaking  Ising transition is expected to survive, which would be consistent with the single specific heat
``peak" seen at the onset  of the spontaneous AHE in Pr$_2$Ir$_2$O$_7$. Monte Carlo studies of the
thermal properties and disorder effects will be discussed elsewhere \cite{lee:J1-J2-pyrochlores}.

The most direct evidence for the scenario advocated here would be a probe which can detect the modulated
Ising order using neutron diffraction.
Landau theory arguments predict that entering such a coexistence phase in a clean system would also lead to a 
weak charge density wave of the Ir electrons at the ordering wavevector of the spiral, from a term 
$\propto \tau^{x,y}(\bs{q}) \rho(-\bs{q})$, and which we find small but finite in our calculations; 
such charge order may be weak 
but can be probed, in principle, using X-ray diffraction. In the presence of an induced charge order, Landau theory arguments
also predict a nonzero $d$-band magnetization, from a term $\propto M \tau^z(\bs{q}) \rho(-\bs{q})$, which is
indeed present as discussed above. Quantum oscillation measurements to detect the ordering-induced
Fermi pockets are desirable, but might be difficult due to the significant AHE and the small size of the pockets.

{\it Conclusion.---} 
In conclusion, we have proposed a mechanism of intrinsic AHE in metallic pyrochlore systems, such as
Pr$_2$Ir$_2$O$_7$,
arising from spiral order of local moments driven by their extended anisotropic RKKY exchange interactions,
and the resulting reconstruction of the electronic band structure to form small Fermi pockets and pairs of Weyl points.
This ordering could occur proximate to an all-in-all-out state of the local moments, and appears to be distinct
from previously proposed spin-chirality scenarios for the AHE in Pr$_2$Ir$_2$O$_7$.

We are grateful to Subhro Bhattacharjee, Hae-Young Kee, Eric Kin-Ho Lee, Jeffrey Rau for useful discussions. 
This work was supported by NSERC, CIFAR, and Center for Quantum Materials at University of Toronto.

\bibliography{AHE-pyrochlores-refs}


\clearpage 

\appendix

\section{Supplementary Material}

\subsection{Local coordinates and primitive vectors in A$_2$B$_2$O$_7$}
First, we define the local coordinates for every A-site pyrochlore lattice. 
The local spin quantization axis ($\hat{z}$) is always pointing towards the center of each tetrahedron
and $\hat{x},\hat{y}$ axes are defined on its basal plane. Table \ref{tab:1} shows
their local coordinates frames depending on the four sublattices of the pyrochlore lattice.
The A-site pseudospin in global coordinates, say $\tilde{\bs{\tau}}_s (\bs{r})$, can be represented as
\be
\tilde{\bs{\tau}}_s (\bs{r}) = \tau_s^x (\bs{r}) \hat{x}_s +  \tau_s^y (\bs{r}) \hat{y}_s + \tau_s^z (\bs{r}) \hat{z}_s 
\ee
where $\bs{\tau}_s (\bs{r}) = \Big( \tau_s^x (\bs{r}) , \tau_s^y (\bs{r}) , \tau_s^z (\bs{r}) \Big)$ 
is the pseudospin at site $\bs{r}$ and sublattice $s$ defined in the local coordinates. 
Both A and B-sites form pyrochlore lattice and each can be viewed as FCC lattice with a four-site basis.
We define the primitive vectors ($b_i$) and the position of four sublattices ($A_i,B_i$) for A and B-sites 
in Table \ref{tab:2}.
\begin{table}[htbp]
\centering
\begin{tabular}{|c| c| c| c| c|}
\hline
$i$ & 0 & 1& 2& 3 \\
\hline \hline
$\hat{x}_i$ & $\frac{1}{\sqrt{2}} (0 1\bar{1} )$ &  $\frac{1}{\sqrt{2}} (0 \bar{1} 1)$ 
& $\frac{1}{\sqrt{2}} (011)$ &  $\frac{1}{\sqrt{2}} (0 \bar{1} \bar{1} )$ \\
$\hat{y}_i$ & $\frac{1}{\sqrt{6}} (\bar{2} 11 )$ &  $\frac{1}{\sqrt{6}} (\bar{2} \bar{1} \bar{1})$ 
& $\frac{1}{\sqrt{6}} (2 1 \bar{1} )$ &  $\frac{1}{\sqrt{6}} (2 \bar{1} 1 )$ \\
$\hat{z}_i$ & $\frac{1}{\sqrt{3}} (111)$ &  $\frac{1}{\sqrt{3}} (1 \bar{1} \bar{1})$ 
& $\frac{1}{\sqrt{3}} (\bar{1} 1 \bar{1})$ &  $\frac{1}{\sqrt{3}} (\bar{1} \bar{1} 1)$\\
\hline 
\end{tabular}
\caption{Local coordinate frames for the four sublattices on the pyrochlore lattice.}
\label{tab:1}
\end{table}
\begin{table}[htbp]
\centering
\begin{tabular}{|c| c| c| c|c|c|}
\hline
$i$ & 0 & 1& 2& 3 \\
\hline \hline
$A_i$ & $(0,1,0)$ &  $(0,\frac{3}{2},\frac{1}{2})$ 
& $(\frac{1}{2},1,\frac{1}{2} )$ &  $(\frac{1}{2},\frac{3}{2},0)$ \\
$B_i$ & $(0,0,0)$ &  $(0,\frac{1}{2},\frac{1}{2})$ 
& $(\frac{1}{2},0,\frac{1}{2})$ &  $(\frac{1}{2},\frac{1}{2},0)$\\
${b}_i$ & &  $(0,1,1)$ & $(1,0,1)$ & $(1,1,0)$ \\
\hline 
\end{tabular}
\caption{Primitive vectors (${b}_i$) and four basis vectors on each A-site ($A_i$) and B-site ($B_i$).}
\label{tab:2}
\end{table}

\subsection{Kondo-like coupling between $A$ and B-sites}
In this section, we derive the explicit form of the Kondo-like coupling between the A-site non-Kramers doublet and 
the B-site Kramers doublet. 
As we argued in the main text, the Ising component of the A-site pseudospin $\tau^z_j$ couples to 
their six neighboring B-site magnetic moments $j_i ^\mu
= c^\dagger_{i \alpha} \frac{\sigma^\mu_{\alpha \beta}}{2}  c_{i \beta}$, 
whereas, the A-site planar components $\tau_j^\pm$ couple to 
the B-site electron density $n_i = c^\dagger_{i \alpha} c_{i \alpha}$. 
Hence, we separate the former and the latter cases and derive the Kondo-like coupling term allowed by lattice symmetry.
First of all, let's consider the former case where $\tau^z_j$ couples to $j_i ^\mu$.
We note that this magnetic coupling term is already derived in Ref.\onlinecite{chen2012magnetic}.
\begin{widetext}
 \begin{eqnarray}
 H_{AB}^{zz} &=&  \sum_{r}
 c_1
    \Big[ \{  j_1^x(r)+j_1^x\left(r-b_2+b_3\right)+j_2^
   y\left(r+b_1-b_2\right)+j_2^y\left(r-b_2+b_3\right)+j
   _3^z(r)+j_3^z\left(r+b_1-b_2\right)   \}
   \tau_0^z(r)
    \nn \\&& \phantom{c_1[}
   + \{  j_0^x\left(r+b_1\right)+j_0^x\left(r+b_
   1-b_2+b_3\right)-j_2^z\left(r+b_1-b_2\right)-j_2^z\left
   (r+b_1-b_2+b_3\right)
   \nn \\&& \phantom{c_1[}
   -j_3^y\left(r+b_1\right)-j_3^y
   \left(r+b_1-b_2\right) \} \tau
   _1^z(r)+  \{ j_0^y\left(r+b_1\right)+j_0^y\left(r+b_
   3\right) 
    \nn \\&& \phantom{c_1[} 
    -j_1^z(r)-j_1^z\left(r+b_3\right)-j_3^x(r)-j_
   3^x\left(r+b_1\right) \} \tau
   _2^z(r) 
   + \{  j_0^z\left(r+b_3\right)+j_0^z\left(r+b_
   1-b_2+b_3\right)
    \nn \\&& \phantom{c_1[}
   -j_1^y\left(r+b_3\right)-j_1^y\left(r
   -b_2+b_3\right)-j_2^x\left(r-b_2+b_3\right)-j_2^x\left(
   r+b_1-b_2+b_3\right)  \}
    \tau _3^z(r) \Big] 
     \nn \\&& \phantom{c_1[}
    +c_2 \Big[
   \{ j_1^y(r)+j_1^y\left(r-b_2+b_3\right)+j_1^
   z(r)+j_1^z\left(r-b_2+b_3\right)+j_2^x\left(r+b_1-b_2
   \right)+j_2^x\left(r-b_2+b_3\right)
    \nn \\&& \phantom{c_1[}
   +j_2^z\left(r+b_1-
   b_2\right)+j_2^z\left(r-b_2+b_3\right)  +j_3^x(r)+j_3^x
   \left(r+b_1-b_2\right)+j_3^y(r)+j_3^y\left(r+b_1-b_2 \right)  \} 
   \tau_0^z(r)
   \nn \\&& \phantom{c_1[}
   + \{ -j_0^y\left(r+b_1\right)-j_0^y\left(r+b
   _1-b_2+b_3\right)-j_0^z\left(r+b_1\right)-j_0^z\left(
   r+b_1-b_2+b_3\right)
   \nn \\&& \phantom{c_1[}
   +j_2^x\left(r+b_1-b_2\right)+j_2^
   x\left(r+b_1-b_2+b_3\right)-j_2^y\left(r+b_1-b_2\right)
   -j_2^y\left(r+b_1-b_2+b_3\right)+j_3^x\left(r+b_1\right)
   \nn \\&& \phantom{c_1[}
   +j_3^x\left(r+b_1-b_2\right)-j_3^z\left(r+b_1\right)
   -j_3^z\left(r+b_1-b_2\right)  \}  \tau
   _1^z(r)
   +\{-j_0^x\left(r+b_1\right)-j_0^x\left(r+b
   _3\right)
   \nn \\&& \phantom{c_1[}
   -j_0^z\left(r+b_1\right)-j_0^z\left(r+b_3\right)
   -j_1^x(r)-j_1^x\left(r+b_3\right)+j_1^y(r)+j_1^y\left(
   r+b_3\right)
   \nn \\&& \phantom{c_1[}
   +j_3^y(r)+j_3^y\left(r+b_1\right)-j_
   3^z(r)-j_3^z\left(r+b_1\right)\} \tau
   _2^z(r)+\{-j_0^x\left(r+b_3\right)-j_0^x\left(r+b
   _1-b_2+b_3\right)
   \nn \\&& \phantom{c_1[}
   -j_0^y\left(r+b_3\right)-j_0^y\left(
   r+b_1-b_2+b_3\right)-j_1^x\left(r+b_3\right)-j_1^x\left(
   r-b_2+b_3\right)+j_1^z\left(r+b_3\right)+j_1^z\left(
   r-b_2+b_3\right)
   \nn \\&& \phantom{c_1[}
 -j_2^y\left(r-b_2+b_3\right)-j_2^y\left(r+b_1-b_2+b_3\right)+j_2^z\left(r-b_2+b_3\right)
   +j_2^z\left(r+b_1-b_2+b_3\right)  \} \tau
   _3^z(r)  \Big]
   \label{eq:6}
 \end{eqnarray}
 \end{widetext}
Here, $\tau_i^z (r)$ is the Ising component of the pseudospin at sublattice $i$ $(=0,1,2,3) $ 
in unit cell labeled by $r$ which is pointing along local $\hat{z}_i$ direction. (See Table \ref{tab:1} 
for their local coordinate frames for the four sublattices on A-site pyrochlore lattice) 
In a similar way, $j_i^\mu ( r )$ is $\mu$ $(=x,y,z)$ component of B-site magnetic moments 
in global cubic coordinates.
Primitive vectors (${ b}_i$) and four basis vectors on each A-site (${ A}_i$)
and B-site (${ B}_i$) are shown in Table \ref{tab:2}. 
There are two independent parameters, $c_1$ and $c_2$ which can not 
be determined by symmetry grounds.
However, two independent parameters can 
be easily understood as follows. For every given site $i$ (on the A-site), there are 
one local $\hat{z}_i$ axis and its basal plane formed by $\hat{x}_i$ and $\hat{y}_i$.(See Table \ref{tab:1}) 
The magnetic coupling can be written as 
$\tau_i^z ( c_j^x j_j^x +  c_j^y j_j^y  + c_j^z j_j^z )$ assuming that $j_j^z$ is aligned along 
their local $\hat{z}_i$ axis. For its basal plane, we have freedom to choose 
$\hat{x}_j$ and $\hat{y}_j$, which leaves $c_j^x = c_j^y$. 
Once the coupling between A-site $i$ and one of
their six neighboring B-site $j$ is determined, 
the coupling with all the other five neighbors are determined 
by lattice $c_3$ rotation and $\sigma_h$ mirror symmetries, 
with two independent parameters $c_j^z$ and $c_j^x=c_j^y$. 

On the other hand, the planar components of the A-site pseudospin are time reversal invariant and 
they do not couple to B-site magnetic moments  but couple to the B-site electron density 
$n_i=c_{i \alpha}^\dagger c_{i \alpha} $. 
In this case, there is only one independent parameter and one can write the coupling term 
as following 
\begin{widetext}
\bea
H_{AB}^\pm &=& c_3  \sum_{r} \Big[
\tau^+_0({r}) \{ \omega ^2
   n_2\left(b_1-b_2+r\right)+\omega ^2
   n_2\left(-b_2+b_3+r\right)+\omega 
   n_3\left(b_1-b_2+r\right) +\omega 
   n_3(r) 
    \nn \\&& \phantom{d}
   +n_1\left(-b_2+b_3+r\right) +n_1(r) \} +\tau^+_1(r) \{ \omega ^2
   n_3\left(b_1+r\right)+\omega ^2 n_3\left(b_1-b_2+r\right)
   \nn \\&& \phantom{c_1[}
   +\omega 
   n_2\left(b_1-b_2+r\right)+\omega 
   n_2\left(b_1-b_2+b_3+r\right)+n_0\left(b_1+r\right)+n_0\left(b_1-b
   _2+b_3+r\right) \} 
     \nn \\&& \phantom{c_1[}
   +\tau^+_2(r) \{ \omega ^2
   n_0\left(b_1+r\right)+\omega ^2 n_0\left(b_3+r\right)+\omega 
   n_1\left(b_3+r\right)+n_3\left(b_1+r\right)+\omega 
   n_1(r)+n_3(r) \}
     \nn \\&& \phantom{d}
   +\tau^+_3(r) \{ \omega ^2
   n_1\left(b_3+r\right)+\omega ^2 n_1\left(-b_2+b_3+r\right)+\omega 
   n_0\left(b_3+r\right)+\omega 
   n_0\left(b_1-b_2+b_3+r\right)
   \nn \\&& \phantom{d}
   +n_2\left(-b_2+b_3+r\right)+n_2\left(
   b_1-b_2+b_3+r\right) \}
   + h.c
\Big]
\label{eq:7}
\eea
\end{widetext}
Extra phases $1, \omega = e^{i 2\pi/3}, \omega^2 $ are present depending on the different 
sublattices due to their local coordinates on the A-site pseudospin.
Combining $ H_{AB}^{zz} $ and $H_{AB}^\pm$, the Kondo-like coupling with three independent 
coupling parameters $c_1,c_2,c_3$ is written as
\be
H_{AB} (c_1,c_2,c_3) = H_{AB}^{zz} (c_1,c_2) + H_{AB}^\pm (c_3).
\label{eq:8}
\ee
In the presence of magnetic ordering on the A-site, 
we take $\bs{\tau}_s ( r) \rightarrow \la \bs{\tau}_s (r) \ra$ which acts as an
effective magnetic field or chemical potential at every B-site itinerant electrons.

\end{document}